\documentstyle[amsfonts,amssymb,amsmath,,epsfig,aps]{revtex} 

\def\sqr#1#2{{\vcenter{\hrule height.#2pt\hbox{\vrule width.#2pt
height#1pt \kern#1pt \vrule width.#2pt}\hrule height.#2pt}}}

\def\hook{\hbox{\vrule height0pt width4pt depth0.3pt
\vrule height7pt width0.3pt depth0.3pt \vrule height0pt width2pt
depth0pt} }
\makeatletter
\def\thepage{1-\@arabic\c@page}
\def\@pnumwidth{2em}
\makeatother

\def\br{\begin{eqnarray}}
\def\er{\end{eqnarray}}
\def\brn{\begin{eqnarray*}}
\def\ern{\end{eqnarray*}}
\def\er{\end{eqnarray}}
\def\beq{\begin{equation}}
\def\eeq{\end{equation}}
\def\vt{\vartheta}
\def\vp{\varphi}
\def\L{{\cal{L}}}

\def\a{\alpha}
\def\b{\beta}
\def\m{\mu}
\def\n{\nu}

\def\d{\delta}

\def\r{\rho}
\def\L{{\mathcal{L}}}
\def\T{{\mathcal{T}}}

\def\C{{\mathcal{C}}}
\def\F{{\mathcal{F}}}

\begin{document}
\title{Conserved currents for general teleparallel models }
\author{Yakov Itin} %\cite{byline}}
\address{Institute of Mathematics,  
Hebrew University of Jerusalem\\
 Givat Ram, Jerusalem 91904, Israel}
\date{\today}
\maketitle
\begin{abstract}
The obstruction for the existence of an energy momentum tensor for  the 
gravitational field is  connected with  differential-geometric features 
of the Riemannian manifold.   
It has not to be valid  for alternative geometrical structures.\\
A teleparallel manifold is defined as a parallelizable
differentiable $4D$-manifold  endowed with a class of smooth coframe fields 
related by global Lorentz, i.e., $SO(1,3)$ transformations.
In this article a general 3-parameter class of teleparallel models 
is considered.
It includes  a 1-parameter subclass of models with the Schwarzschild
coframe solution (generalized teleparallel equivalent of gravity).\\
A new form of the coframe field equation is derived here from the general 
teleparallel Lagrangian
by introducing the notion of a 3-parameter conjugate field strength $\F^a$.
The field equation turns out to have a form completely similar
to the Maxwell field equation $d*\F^a=\T^a$. 
By applying the Noether procedure, the source 3-form  $\T^a$ is shown to be 
connected with the diffeomorphism invariance of the Lagrangian. 
Thus the source of the coframe field 
is interpreted as the total conserved energy-momentum current of the system. \\
A reduction of the conserved current to  the Noether current and the 
Noether charge for the coframe field is provided. 
The energy-momentum tensor  is defined as a map of the module of
current 3-forms into the module of vector fields. 
Thus an energy-momentum tensor for the coframe field is 
defined in a diffeomorphism invariant and a translational covariant way. 
The total energy-momentum current of a system is conserved. 
Thus a redistribution of the energy-momentum current between material 
and coframe (gravity) field is possible in principle, unlike as in GR.\\
The energy-momentum tensor is calculated for various teleparallel
models: the pure Yang-Mills type model, the anti-Yang-Mills type model and
the generalized teleparallel equivalent of GR. 
The latter case can serve as a very close  alternative  
to the GR description of gravity. 
\end{abstract}

%%%%%%%%%%%%%%%%%%%%%%%%%%%%%%%%%%%%%%%%%%%%%%%%%%%%
\section{Introduction}          %%%%%%% 1
%%%%%%%%%%%%%%%%%%%%%%%%%%%%%%%%%%%%%%%%%%%%%%%%%%%%
The concept of an energy-momentum tensor for the gravitational field is, 
undoubtedly, the most puzzling issue in general relativity (GR). 
This quantity is well defined for  other classical fields acting 
in a  fixed  geometrical background and has the following properties. 
It  is  
\begin{itemize}  
\item[(i)]
{\it local} - i.e., constructed only from the fields taken at some point 
on a manifold and the derivatives  of the fields taken at the  same point,
\item[(ii)]
{\it covariant} - i.e.,   transforms as a tensor under  
diffeomorphisms of the manifold,
\item[(iii)] 
{\it conserved} - i.e., it satisfies the covariant divergence equation 
${T^\mu}_{\nu;\mu}=0$, 
\item[(iv)] 
{\it ``the first integral of the field equation''} - 
it is derivable from 
the field equations by integration and includes 
the field derivatives with an  order of one less than 
the order of the field equation.
\end{itemize}
It is well known that in Einstein's theory of gravity a quantity 
satisfying the conditions  listed above does not exist. 
This fact is usually related to the existence of the equivalence principle. 
It implies that the gravitational field can not be detected at a point 
as a covariant object. 
This conclusion can also be  viewed  as a purely differential-geometric fact. 
Indeed, the components of the metric tensor  are managed by a system 
of second order partial differential 
equations.
Thus the energy-momentum
quantity has to be a local tensor constructed from the metric components and 
their first order derivatives. 
The corresponding theorem of  (pseudo) Riemannian geometry 
(due to Weyl \cite{Weyl}) states that  every expression  
of such a type is trivial. 
Thus the objection for the existence of a gravitational 
 energy-momentum tensor is 
directly related to the geometric property of the (pseudo) 
Riemannian manifold.\\
It is natural to expect that  this objection can be lifted 
in an alternative geometric model of gravity.\\
In resent time {\it teleparallel  structures} in the geometry of spacetime 
has evoked  a considerable  interest  for various reasons. 
They was considered as a possible physical relevant geometry by itself 
as well as an essential part of generalized non-Riemannian theories 
such as the Poincar{\'e} gauge theory or metric - affine gravity. 
Another important subject are the various applications of the frame technique 
in physical theories based on classical (pseudo) Riemannian geometry. \\
The construction of a manifold endowed  with a smooth field of frames 
(rep{\'e}re, vierbein) originated in the  ``Rep{\'e}re Mobile'' method 
of Darboux-E. Cartan \cite {Ca} in differential geometry.  
Weitzenb\"{o}ck \cite{We} was the first who recognized that this 
construction can be also 
viewed as a self-sufficient geometrical structure. 
The geometrical structure of the teleparallel manifold was studied 
intensively in the first third of the last century - see \cite{Thomas} 
and the references therein.\\
The teleparallel description of  gravity has been also studied for a 
long time. 
The pioneering works of Cartan \cite {Ca} and Einstein \cite{Ei} 
dealt meaningfully with  various models of unified 
(gravitational-electromagnetic) field theory. \\
Investigations in gauge field theory of gravity and in Einstein-Cartan gravity 
(see  \cite{Hehl4}, \cite {Kawai1} and the references therein), renewed the  
interest in teleparallel geometry. 
In the framework of the general geometrical metric-affine theory of gravity 
(MAG) \cite{{hehl95}}, the teleparallel structure appears as one of 
the basic substructures.
Accordingly, the  general Lagrangian of MAG includes the pure teleparallel 
terms as a separate part. 
For investigations in this area see Refs. \cite{Kop} to \cite{Mielke}.\\
The dynamics of classical fields is completely determined
by an appropriate integral functional of action.
One derives the field equation by applying the least action principle.
Conserved field currents are the first
integrals of the field equation. 
Hence they can be derived, in principle, from the field equations  
by algebraic manipulations. 
However, one cannot describe in this way the origin and the meaning 
of the conservation law.  
The Noether theorem (see e.g. \cite{Bruhat}, \cite{Trautman}) provides
a natural description of conservation laws. 
The conserved currents  are associated with the invariant properties of the
action under certain groups of transformations.
The Noether theorem  also provides 
an algorithmic procedure for the actual construction of a conserved current.\\ 
Recently the classical Noether technique was extended and reformulated in the
 language of modern differential geometric. 
The investigation into the  Noether 
technique by  Wald and his collaborators
\cite{Wald} motivated mostly by the problem of the black hole entropy. 
Important progress in the variational technique was 
provided by the free variation bicomplex of Anderson  \cite{Anderson}.
Results similar to those of Wald were obtained in the 
framework of this general theory.  
In both approaches the general diffeomorphic  Lagrangians was 
studied. \\
In this article we will make use of the covariant Noether
 procedure for specific Lagrangians with additional (gauge) symmetry.
The outline  of the paper is as follows: \\
We start in the first section with Maxwell-scalar system in 
flat Minkowski space. 
The field equations, the conserved currents, and the energy-momentum tensor 
are exhibited by explicitly covariant expressions of differential forms. 
This is used for the comparison with 
the teleparallel models. 
The second section serves as a brief survey of the teleparallel 
description of gravity.\\
Our main results are presented in the third section. 
We consider a coframe-scalar system with the most general 
odd quadratic coframe Lagrangian. 
The field equation is derived in a form almost literally similar 
to the Maxwell equation. 
By applying the Noether procedure, the conserved 
current associated with the diffeomorphism invariance of the 
Lagrangian is derived. 
It is interpreted as the total energy-momentum current of the system. 
This is a conserved current. 
It serves as the source of the coframe field. 
Consequently, a redistribution  of energy between material 
and gravitational (coframe) fields is possible in principle.  \\
The notion of the Noether 
current and the Noether charge for the coframe field are introduced. 
The energy-momentum tensor  is defined as a map of the module of
current 3-forms into the module of vector fields. 
Thus an energy-momentum tensor for the coframe field is 
defined in a diffeomorphism invariant and a translational covariant way.\\
In the fourth section  the energy-momentum tensor is calculated for 
various teleparallel
models: the pure Yang-Mills type model, the anti-Yang-Mills type model, and 
the generalized teleparallel equivalent of gravity. 
The latter case can serve as a very close alternative 
to the GR description of gravity. 
%%%%%%%%%%%%%%%%%%%%%%%%%%%%%%%%%%%%%%%%%%%%%%%%%%%%
\section{Electromagnetic-scalar system}  %%%% 2
%%%%%%%%%%%%%%%%%%%%%%%%%%%%%%%%%%%%%%%%%%%%%%%%%%%%
Before diving into the consideration of the teleparallel models, let us start 
  with a rather simple example of a system that includes  
a gauge vector field   and a scalar field. 
We consider this example  mostly in order to give a brief account of the
Noether procedure and to fix the notations employed in this paper. 
It should be noted, however, that the  teleparallel description of 
gravity will be formulated in the next section 
in a form  very similar to the familiar 
description of the electromagnetic field.  
A coframe field (i.e.  a set of four 1-forms)  is a  basic 
dynamical variable in teleparallel 
gravity,  while   electromagnetic potential 1-form serves as  
the basic variable in Maxwell theory. 
As a  consequence of that, the field equations, the 
conserved currents and the energy-momentum tensor 
in both theories will turn out to be  of similar form.  
%%%%%%%%%%%%%%%%%%%%%%%%%%%%%%%%%%%%%%%%%%%%%%%%%%%%
\subsection{Lagrangian}                  %%%%%%% 2.1
%%%%%%%%%%%%%%%%%%%%%%%%%%%%%%%%%%%%%%%%%%%%%%%%%%%%
Let  the Minkowski space of special relativity be given, i.e., 
a flat $4D$-manifold $M$ with a metric
$\eta_{ab}=(-1,+1,+1,+1)$ and with vanishing torsion.
Consider  a real even 1-form field  
$A$ which couples minimally to  a complex even scalar field $\vp$.
The total Lagrangian density for the system
is given by an odd  differential 4-form
\begin{equation}\label{2.1}
L={}^{(e)}L+{}^{(s)}L=\frac 12 F\wedge*F-
\frac 12  D\vp \wedge*\overline{D\vp},
\end{equation}
where 
\begin{equation}\label{2.2}
F=dA, \qquad \textrm{and} \qquad D\vp=(d+ieA)\vp
\end{equation}
are the even forms of the corresponding field strengths. 
$\overline{D\vp}$ means the complex conjugate, and 
$*$ denotes the Hodge dual map defined 
by the flat metric $\eta_{ab}$. 
Our conventions for the operations on forms are listed in  the Appendix.\\
The Lagrangian density (\ref{2.1}) is invariant under the actions of 
two continuous groups: the Poincar{\'e} group of 
transformations of M, and
the internal $U(1)$-gauge group of transformations of the field $A$.
Hence   two conserved
currents of a different nature is expected  for the system (\ref{2.1}).\\
The infinitesimal variation of the Lagrangian  is given by the
variational relation 
\begin{equation}\label{2.3}
\d L= \d F\wedge*F - {\rm Re}\Big(\d(D\vp)\wedge *\overline{D\vp}\Big).
\end{equation}
Applying the Leibniz rule to extract  the total derivatives and using 
(\ref{2.2}) we obtain an equivalent form of the variation relation (\ref{2.3})
\br\label{2.4}
\d L&=&\d A\wedge \Big(d*F+e {\rm Im}(\vp*\overline{D\vp})\Big)+
{\rm Re}(\d\vp\overline{D*D\vp})\nonumber\\
&&+d\Big(\d A\wedge*F-{\rm Re}(\d\vp*\overline{D\vp})\Big).
\er
%%%%%%%%%%%%%%%%%%%%%%%%%%%%%%%%%%%%%%%%%%%%%%%%%%%%
\subsection{Field equations}      %%%%%%% 2.2
%%%%%%%%%%%%%%%%%%%%%%%%%%%%%%%%%%%%%%%%%%%%%%%%%%%%
The variation (\ref{2.4}) of the Lagrangian  has to vanish for a class 
of arbitrary infinitesimal variations of the dynamical variables 
$A$ and $\vp$, hence it should be zero 
also  for a subclass of independent variations 
vanishing at infinity. 
The total derivative term in (\ref{2.4}) vanishes for such variations, 
while the first two terms  result in two field equations. 
The field equation for the scalar field is
\begin{equation}\label{2.5}
 D*D\vp=0,
\end{equation}
while the field equation for the electromagnetic field is
\begin{equation}\label{2.6}
d*F={}^{(s)}I, \qquad {\mathrm{or}} \qquad  d*dA={}^{(s)}I.
\end{equation}
The odd 3-form in the right hand side of (\ref{2.6}) is
the  electromagnetic current produced by  the complex scalar field 
\begin{equation}\label{2.7}
{}^{(s)}I=-e{\rm Im}(\vp*\overline{D\vp}),
\end{equation}
which serves as a source for the field strength $F$.\\
Observe  the structure of the field equations 
(\ref{2.6}). 
The left hand side is the exterior derivative 
of the Hodge dual of the odd strength (2-form) while the right hand side is 
the odd form of a current (3-form).
We will see later that this  is a generic structure. \\
An immediate consequence of the field equation (\ref{2.6}) is the 
conservation law for the electromagnetic current 
\begin{equation}\label{2.8}
d \ {}^{(s)}I=0
\end{equation}
This current is known to be associated with the
$U(1)$-gauge  symmetry of the Lagrangian (\ref{2.1}).
Let us turn, however, to another conserved current associated with the
invariance of the total Lagrangian (\ref{2.1}) under the  action of
Poincar{\'e} group of the space-time transformations.
%%%%%%%%%%%%%%%%%%%%%%%%%%%%%%%%%%%%%%%%%%%%%%%%%%%%
\subsection{Conserved currents}         %%%%%%% 2.3
%%%%%%%%%%%%%%%%%%%%%%%%%%%%%%%%%%%%%%%%%%%%%%%%%%%%
On shell, i.e., for the fields satisfying the field equations (\ref{2.5}) and 
(\ref{2.6}), the variation relation (\ref{2.4}) reduces to 
\begin{equation}\label{2.9}
d\Big({}^{(e)}\Theta+{}^{(s)}\Theta\Big)-\d L=0,
\end{equation}
where the 3-forms for the electromagnetic and for 
the  scalar fields  are
\begin{equation}\label{2.10}
{}^{(e)}\Theta=\d A\wedge*F, \qquad \textrm{and} \qquad 
{}^{(s)}\Theta=-{\rm Re}(\d\vp*\overline{D\vp}).
\end{equation}
Consider  a vector field  $X$ which represents an infinitesimal
action of the Poincar{\'e} group on $M$.
Let the variations of the fields
$\d A$ and $\d\vp$ be produced by the Lie derivative operator $\L_X$ taken 
with respect  to $X$.
Since the Lagrangian (\ref{2.1}) is Poincar{\'e} invariant, its variation, 
being generated by the variations of the fields,
will also be  produced by the Lie derivative $\L_X$. 
Thus the equations  
$\d\vp=\L_X\vp,\quad \d A=\L_X A $
result in
$\d L=\L_X L.$  
We will make  use of a known formula for the Lie derivative of an
arbitrary $p$-form $\a$
\begin{equation}\label{2.13}
\L_X \a=d(X\hook \a)+X\hook d\a.
\end{equation}
Note that the first term in the right hand side of (\ref{2.13}) vanishes for 
scalars. The second term is zero for $4$-forms.
Hence, the Lie derivative and consequently the variation of an 
arbitrary Lagrangian 4-form is a total derivative. 
After  substituting the corresponding expressions for the 
variations into (\ref{2.9}) it takes  a form of a conservation
law for a certain odd 3-form $J(X)$
\begin{equation}\label{2.14}
dJ(X)=0.
\end{equation}
This  3-form  is expressed as the sum of the electromagnetic and 
the scalar parts
\begin{equation}\label{2.15}
J(X)={}^{(e)}J(X)+{}^{(s)}J(X).
\end{equation}
This fact is in agreement with the minimal coupling form of the Lagrangian 
(\ref{2.1}). 
The  explicit expressions for the currents are 
\br\label{2.16}
{}^{(e)}J(X)&=&\Big(d(X\hook A)+(X\hook F)\Big)\wedge*F-X\hook {}^{(e)}L,\\
\label{2.17}
{}^{(s)}J(X)&=&-{\rm Re}\Big((X\hook d\vp)*\overline{D\vp}\Big)-
X\hook {}^{(s)}L.
\er
The conserved 3-form $J(X)$ can be considered as a preliminary
candidate  for the total energy-momentum current for the
scalar-electromagnetic system. 
It is odd, covariant, local, and associated with the transformations 
of the manifold. 
Unfortunately, the electromagnetic 3-form
(\ref{2.16}) includes a term (the first  one) with a non-algebraic
dependence on the vector field $X$.
Consequence will be shown that only a current which depends 
linearly (algebraic) on an arbitrary vector field admits a 
reformulation in terms of the energy-momentum tensor.
Observe also another problem with the  reduction (\ref{2.15}).  
The  electromagnetic  part (\ref{2.16}) 
and the scalar  part (\ref{2.17}) are  not separately gauge invariant.\\
The 3-form $J(X)$ can  be amended in order to avoid the non-algebraic 
dependence on an arbitrary vector $X$ and also  to recover the separate 
gauge invariance of its pieces. 
For that, we use the Leibniz rule for extracting the total derivative in the 
first term   of (\ref{2.16}).  
Apply  the field equation (\ref{2.6}) 
to obtain a new reduction 
\begin{equation}\label{2.18}
J(X)={}^{(e)}\T (X)+{}^{(s)}\T (X)+d\Big({}^{(e)}Q(X)\Big),
\end{equation}
where the  3-forms are 
\br\label{2.19}
{}^{(e)}\T(X)&=&(X\hook F)\wedge *F-\frac 12 X\hook (F\wedge *F),\\
\label{2.20}
{}^{(s)}\T(X)&=&-{\rm Re}\Big((X\hook D\vp)*\overline{D\vp}\Big)
+\frac 12 X\hook (D\vp\wedge *\overline{D\vp}).
\er
As for  the 2-form ${}^{(e)}Q(X)$, it is exhibited as 
\begin{equation}\label{2.21}
{}^{(e)}Q(X)=(X\hook A)*F.
\end{equation}
The 3-forms ${}^{(s)}\T(X)$ and ${}^{(e)}\T(X)$ are odd, covariant,
local and their sum 
\begin{equation}\label{2.22}  
\T(X)={}^{(e)}\T (X)+{}^{(s)}\T (X)
\end{equation}
is conserved:
\begin{equation}\label{2.23}   
d\T(X)=0.
\end{equation}
Thus $\T(X)$ can be interpreted as the {\it total conserved current}
 of the system.
Accordingly, by \cite{Wald}  the  3-forms
(\ref{2.19}), (\ref{2.20}) will be refereed to as 
{\it the electromagnetic and the scalar Noether currents} correspondingly. 
As for the odd 2-form $Q(X)$ it can be identified with the
{\it Noether charge} \cite{Wald}. 
Note that this object is not  gauge invariant and depends only 
on the electromagnetic field variables.\\
As for the gauge invariance of the reduction (\ref{2.15}),  
it is recovered: the  
electromagnetic part (\ref{2.19}) and the scalar part (\ref{2.20}) 
are both  gauge invariant.\\
Although, the 3-forms ${}^{(s)}\T(X)$ and ${}^{(m)}\T(X)$   
depend on an arbitrary vector field $X$, this dependence is algebraically
linear.
Thus every one of these currents can
be treated as  a linear map of the module
of vector fields on $M$ into the module of 3-forms on $M$, 
i.e., as  a tensor field.
A natural representation of a linear map on vector spaces  (on modules)  
is obtained by its action on the basis vectors.
In this case the currents are refereed to as {\it the canonical currents}
of the corresponding fields.
Take $e_a=\partial/\partial x^a$ to obtain 
{\it the canonical energy-momentum current of the electromagnetic field}
\begin{equation}\label{2.24}   
{}^{(e)}\T_a:={}^{(e)}\T(e_a)=(e_a\hook F)\wedge *F-
\frac 12 e_a\hook (F\wedge *F),
\end{equation}
and {\it the canonical energy-momentum current of the scalar field}
\begin{equation}\label{2.25}   %s-cur3}
{}^{(s)}\T_a:={}^{(s)}\T(e_a)=
-{\rm Re}\Big((e_a\hook D\vp)*\overline{D\vp}\Big)+
\frac 12 e_a\hook (D\vp\wedge *\overline{D\vp}).
\end{equation}
The {\it canonical Noether charge} 2-form for the electromagnetic
field can also be defined: 
\begin{equation}\label{2.26}    %em-2curr2}
{}^{(e)}Q_a:={}^{(e)}Q(e_a)=(e_a\hook A)*F.
\end{equation}
Recall that this quantity is not  gauge invariant.
%%%%%%%%%%%%%%%%%%%%%%%%%%%%%%%%%%%%%%%%%%%%%%%%%%%%
\subsection{The energy-momentum tensor}     %%%%%%% 2.5
%%%%%%%%%%%%%%%%%%%%%%%%%%%%%%%%%%%%%%%%%%%%%%%%%%%%
Let us introduce now the notion of the energy-momentum tensor 
by this differential-form formalism. 
We are looking for a second rank tensor field of a type $(0,2)$.
Such  a tensor can always be treated as a  bilinear map
$$T: \ {\mathcal X(M)}\times{\mathcal X(M)} \to {\mathcal F(M)},$$
where ${\mathcal F(M)}$ is the algebra of $C^\infty$-functions on $M$
while ${\mathcal X(M)}$ is the ${\mathcal F(M)}$-module of vector fields
on $M$. 
The unique way to construct a scalar from a 3-form and a vector is 
is to take the Hodge dual of the 3-form and to 
contract the result by the vector. 
Consequently, we define the energy-momentum tensor as 
\begin{equation}\label{2.27}  %tem1}
T(X,Y):=Y\hook *\T(X).
\end{equation}
Observe that this quantity is a tensor if and only if the 3-form current
$\T$ depends  linearly (algebraic) on the vector field $X$.
Certainly, $T(X,Y)$ is not symmetric in general.  
The canonical form of the energy-momentum  tensor is defined as 
\begin{equation}\label{2.28}    %tem3}
T_{ab}:=T(e_a,e_b)=e_b\hook *\T_a.
\end{equation}
Another useful form of this tensor can be obtained from (\ref{2.28}) 
by applying the rule (\ref{A.15}) 
\begin{equation}\label{2.29}
T_{ab}=-*(\vt_b\wedge \T_a).
\end{equation}
The familiar relation ${T_a}^b=\eta^{bc}T_{ac}$ defines tensors of a
type $(1,1)$
\begin{equation}\label{2.30}
{T_a}^b=-*(\vt^b\wedge \T_a),
\end{equation}
and
\begin{equation}\label{2.31}
{T^a}_{b}=-*(\vt_b\wedge \T^a),
\end{equation}
which are different for non-symmetric $T_{ab}$.  
By applying the rule (\ref{A.9}) the relation (\ref{2.28}) 
can be converted into
\begin{equation}\label{2.32}
\T_a={T_a}^b*\vt_b.
\end{equation}
Thus the components of the energy-momentum tensor are 
the coefficients of the current $\T_a$ in the dual basis
$*\vt^a$ of the vector space $\Omega^3$ of odd 3-forms.\\
In order to justify our definition of the energy-momentum tensor
let us first   show that the current satisfies the ordinary conservation law. 
The coframe $\vt^a=0$ is close, thus $d*\vt_b=0$. 
From (\ref{2.32}) we derive 
$$d\T_a=d{T_a}^b\wedge *\vt_b  =-{{T_a}^b}_{,b}*1.$$
Hence the differential-form conservation law $d\T_a=0$
is equivalent to the tensorial conservation law ${{T_a}^b}_{,b}=0$ and 
conversely.\\ 
Let us write down  the explicit forms  of the   energy-momentum tensor.\\
{\bf For the electromagnetic field} we obtain by substituting (\ref{2.19}) in
(\ref{2.26}) and using (\ref{A.15},\ref{A.17})
\begin{equation}\label{2.33}       
{}^{(e)}T(X,Y)=-*\Big((X\hook F)\wedge *(Y\hook F)\Big)-<X,Y>* \ L.
\end{equation}
By this formula the energy-momentum tensor in an explicitly symmetric form: 
\begin{equation}\label{2.34}        
{}^{(e)}T(X,Y)={}^{(e)}T(Y,X).
\end{equation} 
The canonical form of the electromagnetic energy-momentum tensor is 
\begin{equation}\label{2.34a}       
T_{ab}={}^{(e)}T(e_a,e_b)=-*\Big((e_a\hook F)\wedge *(e_b\hook F)\Big)-\eta_{ab}* \ L.
\end{equation}
In a specific coordinate chart $\{x^\mu\}$ we take the coordinate 
basis vectors 
$X=\partial_\a$ 
and $Y=\partial_\b$
 to obtain the familiar  expression for the energy-momentum tensor of 
the electromagnetic field 
 \begin{equation}\label{2.35}  %%Check the sign ????
{}^{(e)}T_{\a\b}:={}^{(e)}T(\partial_\a,\partial_\b)=
 -F_{\a\m}{F_\b}^\m+\frac 14 \eta_{\a\b}F_{\m\n}F^{\m\n}.
\end{equation}
This tensor is obviously traceless: $ \ \eta ^{\a\b}T_{\a\b}=0$.\\
{\bf For the scalar field} we obtain using (\ref{2.20}) an explicitly 
symmetric form of the energy-momentum tensor
\begin{equation}\label{2.36} 
{}^{(s)}T(X,Y)=-{\rm Re}\Big((X\hook D\vp)*(Y\hook \overline{D\vp})\Big)-<X,Y> *L.
\end{equation}
In a specific coordinate chart $\{x^\mu\}$ we use 
the covariant derivative 
notations $D\vp=D_\a \vp dx^\a$ 
to derive the familiar expression for the  energy-momentum
tensor of a complex scalar field.
 \begin{equation}\label{2.37}    %%Check the sign ????
{}^{(s)}T_{\a\b}=-{\rm Re}(D_\a\vp D_\b\overline \vp)+
\frac 12 \eta_{\a\b}(D_\mu\vp\overline {D^\mu\vp}).
\end{equation}
%%%%%%%%%%%%%%%%%%%%%%%%%%%%%%%%%%%%%%%%%%%%%%%%%%%%
\section{Teleparallel gravity}            %%%%%% 3
%%%%%%%%%%%%%%%%%%%%%%%%%%%%%%%%%%%%%%%%%%%%%%%%%%%%
Let us give a brief account of gravity on teleparallel manifolds.  
Consider a coframe field $\{\vt^a, \ a=0,1,2,3\}$ defined 
on a $4D$  differential manifold $M$.
The 1-forms $\vt^a$ declared to be pseudo-orthonormal.
This determines completely  a metric on the manifold $M$
via the relation 
\begin{equation}\label{3.1} 
g=\eta_{ab}\vt^a\otimes\vt^b.
\end{equation}
Thus it is possible to consider the coframe field $\vt^a$ as
a basic dynamical variable  and to treat  the metric $g$ as only
a secondary structure.\\ 
In order to have an isotropic structure on $M$ (without peculiar directions)
the coframe variable have to  be defined only up to
{\it global pseudo-rotations} of the group  $SO(1,3)$.
Thus, the truly dynamical variable is  the equivalence
class of coframes $[\vt^a]$ is , while 
the global pseudo-rotations produce  the  equivalence relation on this class. 
Hence, in addition to the invariance under the diffeomorphic transformations
of the manifold $M$, the basic geometric structure
has to be  global $SO(1,3)$ invariant.\\
Recall the well known property of the teleparallel geometry: 
it is possible to define the parallelism of two vectors
at different points by 
comparing the components of the vectors in local frames. 
Namely, two vectors (1-forms) are parallel if the corresponding
components are proportional when 
referred to a local frame (coframe). 
This {\it absolute  parallelism} can be treated as a global
path independent parallel transport.
In the affine-connections formalism such a transport is described 
by existence of a special
teleparallel connections of vanishing curvature \cite {hehl95}.
The Riemannian curvature of the manifold 
which is constructed from the metric (\ref{3.1}) is non-zero, in general. \\
Gravity is described by the teleparallel geometry
in a way similar to Einstein theory, i.e.,  by differential-geometric 
invariants of the structure.\\
Looking for such invariants,  an important distinction between 
the metric and the teleparallel structures emerges.\\
{\it The metric structure} admits diffeomorphic invariants of the second 
order or greater. 
The metric invariants of the first order are trivial. 
The unique invariant of the second order is the scalar curvature. 
This expression plays the 
role of an  integrand in the Einstein-Hilbert action.\\
{\it The teleparallel structure} admits diffeomorphic and $SO(1,3)$ global 
invariants of the first order. 
The simplest example of such invariants is the expression $e_a\hook d\vt^a$. 
The class of quadratic diffeomorphic invariant and global 
covariant operators is exhibited in \cite{i-k}. 
These operators  serve to construct  a general  class of field equations \cite{i-k}. 
Restrict the consideration  to field equations derivable 
from an action integral. The corresponding Lagrangian has to be   
odd, quadratic (in the first order derivatives of the coframe field 
$\vt^a$),  diffeomorphic, and global $SO(1,3)$ invariant.  
 It can be constructed as a linear combination of three 
Weitzenb\"{o}ck  quadratic  teleparallel invariants \cite{We}.  
The symmetric form of this Lagrangian  is \cite{Hehl98}
\begin{equation}\label{3.2}         
L=\frac 12 \sum_{i=1}^3 \rho_{i} \; {}^{(i)}L\
\end{equation}
with 
\br\label{3.3}   
{}^{(1)}L &=&d\vt^a \wedge *d\vt_a,\\
\label{3.4}   
{}^{(2)}L &=&
\Big(d\vt_a \wedge \vt^a \Big) \wedge*\Big(d\vt_b\wedge\vt^b\Big), \\
\label{3.5}     
{}^{(3)}L &=& 
(d\vt_a \wedge\vt^b ) \wedge *\Big(d\vt_b \wedge \vt^a \Big).
\er 
The coefficients $\r_i$ are dimensionless free parameters of the theory.\\
Every term of the Lagrangian (\ref{3.2}) is independent of  
a specific choice of a coordinate system and invariant under the global 
(rigid) $SO(1,3)$ transformations of the coframe. 
Thus different choices of the free parameters $\r_i$ yield 
different translation invariant classical field models. 
Some of them are known to  be applicable 
 for description  of the gravitational field.\\
The field equation is derived from the Lagrangian (\ref{3.2}) 
in the form \cite{Kop},\cite{Hehl98}
\br\label{3.6}    
&&\!\!\!\!\!\!\!\!\!\!\!\!\!\!\!\!\rho_1\Big(
2d*d\vt_a+
e_a\hook(d\vt^b\wedge*d\vt_b)-
2(e_a\hook d\vt^b)\wedge*d\vt_b\Big)+\nonumber\\
&&\!\!\!\!\!\!\!\!\!\!\!\!\!\!\!\!\rho_2\Big(
-2\vt_a \wedge d *(d\vt^b\wedge \vt_b)+
2d\vt_a \wedge * ( d\vt^b \wedge \vt_b)+\nonumber\\
&&\!\!\!\!\!
e_a\hook\Big(d\vt^c\wedge\vt_c\wedge*(d\vt^b\wedge\vt_b)\Big)-
2(e_a\hook d\vt^b)\wedge\vt_b\wedge*(d\vt^c\wedge\vt_c)
\Big)+\nonumber\\
&&\!\!\!\!\!\!\!\!\!\!\!\!\!\!\!\!\rho_3\Big(
-2\vt_b \wedge d*( \vt_a \wedge d \vt^b )+
2d\vt_b\wedge*(\vt_a\wedge d\vt^b)+\nonumber\\
&&\!\!\!\!\!
e_a\hook\Big(\vt_c\wedge d\vt^b\wedge*(d\vt^c\wedge\vt_b)\Big)-
2(e_a\hook d\vt^b)\wedge\vt_c\wedge*(d\vt^c\wedge\vt_b )
\Big)=0.
\er
In \cite{it5} the  class of ``diagonal'' 
spherical-symmetric static  solution to the field equation
(\ref{3.6}) is described. 
It was also shown there that a solution 
with Newtonian behavior at infinity  appears only in the case $\rho_1=0.$
This is the Schwarzschild  coframe 
\br\label{3.8}   
\vt^0&=&\frac{1-m/{2r}}{1+ m/{2r}}dt,\qquad \vt^i=
\Big(1+\frac m{2r}\Big)^2dx^i, \qquad i=1,2,3,
\er
which yields via (\ref{3.1}) the Schwarzschild metric in 
isotropic coordinates. \\ 
Few remarks on the structure of the field equation (\ref{3.6}) 
are now in order. \\
On one hand, the coframe field is a complex of $16$ independent variables 
at every point of $M$ while the symmetric metric tensor field has only 
$10$ independent components. 
An additional requirement of {\it local} $SO(1,3)$ {\it invariance}, 
which satisfied in the case 
\begin{equation}\label{3.9}   
\rho_1=0, \qquad \rho_2+2\rho_3=0
\end {equation} 
 restricts the set of 16 independent variables to a subset of $10$ variables.  
This subset is certainly related  to 10  independent components of the metric. \\
On the second hand for an arbitrary choice  of the parameters $\rho_i$ 
the field equation (\ref{3.6}) is a system of 16 independent equations. 
This system can be  
reduced to two covariant systems - a symmetric tensorial system of 
10 independent equations and an antisymmetric tensorial system of 6 
independent equations.  
In {\cite{it4} it was shown that, if 
  (\ref{3.9}) holds, the antisymmetric equation 
vanishes identically.  
Thus in the case of local $SO(1,3)$ invariance  
the system  (\ref{3.6}) is restricted to a system of 10 
independent equations for 10 independent variables. 
Therefore 
 the coframe (teleparallel) structure coincides with 
the metric structure. 
The local $SO(1,3)$ invariant teleparallel model  is referred to as the 
{\it teleparallel equivalent of gravity} \cite {Mielke}-\cite{Per}.
%%%%%%%%%%%%%%%%%%%%%%%%%%%%%%%%%%%%%%%%%%%%%%%%%%%%
\section{Coframe-scalar system}             %%%%%%% 4
%%%%%%%%%%%%%%%%%%%%%%%%%%%%%%%%%%%%%%%%%%%%%%%%%%%%
%%%%%%%%%%%%%%%%%%%%%%%%%%%%%%%%%%%%%%%%%%%%%%%%%%%%
\subsection{The total Lagrangian}               %%%%%%% 4
%%%%%%%%%%%%%%%%%%%%%%%%%%%%%%%%%%%%%%%%%%%%%%%%%%%%
Let  a differential  $4D$-manifold $M$ be given.
Consider a system containing  two smooth fields defined on $M$: 
an even  coframe field $\vt^a$ 
and an even (real) scalar field $\vp$. 
Our goal is to derive a conserved current expression 
for the coframe  field for a set 
of models parameterized by the constants $\rho_i$. 
As in the case of the electromagnetic field, the scalar field $\vp$ will play 
a role of an indicator of a true current.\\
The minimal coupling total Lagrangian 
density of the system is an odd 4-form, which 
can be written as a sum of the 
coframe Lagrangians with  the scalar Lagrangian
\begin{equation}\label{4.1}   
L={}^{(c)}L+{}^{(s)}L=\frac{1}{2}\sum_{i=1}^3 \rho_{i} \; {}^{(i)}L-
\frac{1}{2}d\vp\wedge *d\vp,
\end{equation}
where the coframe Lagrangians  ${}^{(i)}L$ are as defined in 
(\ref{3.3}-\ref{3.5}).\\
The standard computations of the variation of a Lagrangian defined on 
a teleparallel manifold are rather complicated. 
It is because one needs to vary not only  the coframe $\vt^a$ 
itself, but also the the dual frame $e_a$ and even the  
Hodge dual operator, 
that  depends on the coframe implicitly \cite{Hehl98}.\\
In order to avoid these technical problems we will rewrite  
the total Lagrangian (\ref{4.1}) in a 
compact form which will be useful for the  
variation procedure.\\
Consider the exterior differentials of the basis 1-forms $d\vt^a$ and 
introduce the $C$-coefficients of their  expansion in the basis of 
even 2-forms $\vt^{ab}$ 
(we use here and later  the abbreviation 
$\vt^{ab\cdots}=\vt^a\wedge \vt^b\wedge \cdots$)
\begin{equation}\label{4.2} 
d\vt^a=\vt^a_{\b,\a}dx^\a\wedge dx^\b:=\frac 12 {C^a}_{bc}\vt^{bc}.
\end{equation}
By definition, the coefficients ${C^a}_{bc}$ are  antisymmetric: 
${C^a}_{bc}=-{C^a}_{cb}.$ 
The explicit expression can be derived from the definition 
(\ref{4.2})
\begin{equation}\label{4.3} 
{C^a}_{bc}:=e_c\hook(e_b\hook d\vt^a).
\end{equation}
In terms of the $C$-coefficients the teleparallel parts of 
the Lagrangian (\ref{4.1})  are 
\br\label{4.4}   
{}^{(1)}L&=& \frac 12 C_{abc}C^{abc}*1,\nonumber\\
{}^{(2)}L&=& \frac 12 C_{abc}\Big(C^{abc}+C^{bca}+C^{cab}\Big)*1,\nonumber\\
{}^{(3)}L&=&\frac 12 \Big(C_{abc}C^{abc}-2{C^a}_{ac}{C_b}^{bc}\Big)*1.         
\er
The form (\ref{4.4}) is useful for a proof of the completeness of the set of 
quadratic invariants \cite{i-k}. 
It is enough to consider all the possible combinations of the indices. 
Thus a linear combination of the Lagrangians (\ref{4.4}) is the most general 
quadratic coframe Lagrangian.  \\
Using  (\ref{4.4}) we rewrite the  coframe Lagrangian 
in a compact form
\begin{equation}\label{4.5}  
{}^{(c)}L=\frac 14 C_{abc}C_{ijk}\lambda^{abcijk}*1,
\end{equation}
where the  constant symbols
\br\label{4.6}  
\lambda^{abcijk}&:=&(\rho_1+\rho_2+\rho_3)\eta^{ai}\eta^{bj}\eta^{ck}+
\rho_2(\eta^{aj}\eta^{bk}\eta^{ci}+\eta^{ak}\eta^{bi}\eta^{cj})\nonumber \\
&&-2\rho_3\eta^{ac}\eta^{ik}\eta^{bj}
\er
are introduced.  
It can be checked, by straightforward calculation, that these $\lambda$-symbols 
are invariant under a 
transposition of the triplets of indices
\begin{equation}\label{4.7}
\lambda^{abcijk}=\lambda^{ijkabc}.
\end{equation}
We also  introduce an abbreviated notation 
 \begin{equation}\label{4.8}
F^{abc}:=\lambda^{abcijk}C_{ijk}.
\end{equation}
As for the scalar Lagrangian, it can also be rewritten in 
a product form similar to 
(\ref{4.5}) by  introducing the scalar notations for the derivatives  
\begin{equation}\label{4.9}  
d\vp=\vp_a\vt^a, \qquad \textrm{or} \qquad \vp_a=e_a\hook d\vp.
\end{equation}
The total Lagrangian (\ref{4.1}) reads now as
\begin{equation}\label{4.10}
L=\frac 14 \Big(C_{abc}F^{abc}-2\vp_a\vp^a\Big)*1.
\end{equation}
%%%%%%%%%%%%%%%%%%%%%%%%%%%%%%%%%%%%%%%%%%%%%%%%%%%%
 \subsection{Variation of the Lagrangian}                %%%%%%% 4
%%%%%%%%%%%%%%%%%%%%%%%%%%%%%%%%%%%%%%%%%%%%%%%%%%%%
Using the symmetry property of the $\lambda$-symbols (\ref{4.7}) 
the variation of the 
total Lagrangian  (\ref{4.10}) takes the form 
\begin{equation}\label{4.11}  
\d L=\frac 12\Big(\d C_{abc}F^{abc}-2\d\vp_a\vp^a\Big)*1-L*\d(*1).
\end{equation}
The variation of the volume element is 
\brn   %\label{4.12} 
\d(*1)&=&-\d(\vt^{0123})=-\d\vt^0\wedge \vt^{123}-\cdots=
-\d\vt^0\wedge *\vt^0-\cdots \nonumber \\
%\label{4.12a} 
&=&\d \vt^m\wedge*\vt_m.
\ern   
Thus
\begin{equation}\label{4.12}
L*\d(*1)=(\d\vt^m\wedge *\vt_m)*L=-\d\vt^m\wedge(e_m\hook L)
\end{equation}
As for the variation of the $C$-coefficients, we calculate them by equating  
the variations  of the two sides of the equation (\ref{4.2})
$$
\d d\vt_{a}=\frac 12 \d C_{amn}\vt^{mn}+C_{amn}\d\vt^{m}\wedge\vt^n.
$$
Using the formulas (\ref{A.12}) and (\ref{A.15}) we derive  
\brn
\d d\vt_{a}\wedge*\vt_{bc}&=&
\frac 12 \d C_{amn}\vt^{mn}\wedge*\vt_{bc}+
C_{amn}\d\vt^{m}\wedge\vt^n\wedge*\vt_{bc}\\
%%
%&=&\frac 12 \d C_{amn}\vt^{m}\wedge *^2(\vt^n\wedge*\vt_{bc})+
%C_{amn}\d\vt^{m}\wedge*^2(\vt^n\wedge*\vt_{bc})\\
%%
&=&-\frac 12 \d C_{amn}\vt^{m}\wedge *(e^n\hook\vt_{bc})-
C_{amn}\d\vt^{m}\wedge*(e^n\hook\vt_{bc})\\
%%
%&=&-\frac 12 \d C_{amb}\vt^{m}\wedge *\vt_{c}+
%\frac 12 \d C_{amc}\vt^{m}\wedge *\vt_{b}\\
%&&-C_{amb}\d\vt^{m}\wedge*\vt_c+C_{amc}\d\vt^{m}\wedge*\vt_b\\
%%
%&=&\frac 12 \d C_{amb}*^2(\vt^{m}\wedge *\vt_{c})-
%\frac 12 \d C_{amc}*^2(\vt^{m}\wedge *\vt_{b})\\
%&&-C_{amb}\d\vt^{m}\wedge*\vt_c+C_{amc}\d\vt^{m}\wedge*\vt_b\\
%%
&=&\d C_{abc}*1-\d\vt^{m}\wedge\Big(C_{amb}*\vt_c-C_{amc}*\vt_b\Big).
\ern
Therefore 
\begin{equation}\label{4.13}
\d C_{abc}*1=\d (d\vt_a)\wedge *\vt_{bc}
+\d\vt^m \wedge (C_{amb}*\vt_c-C_{amc}*\vt_b).
\end{equation}
The variation of the scalar field coefficients $\vp_a$ can  also 
be calculated using their definition (\ref{4.9}). 
We write 
$$\d(d\vp)=\d\vp_a\vt^a+\vp_a\d\vt^a,$$
consequently, 
\begin{equation}\label{4.14}
\d\vp^a*1=\d(d\vp)\wedge*\vt_a-\vp_m\d\vt^m\wedge*\vt_a.
\end{equation}
After substituting (\ref{4.12}--\ref{4.14}) into (\ref{4.11})
the variation of the Lagrangian  takes the form
\brn
\d L&=&
\frac 12 F^{abc}\Big(\d (d\vt_{a})\wedge*\vt_{bc}+\d\vt^m\wedge(C_{amb}*\vt_c-
C_{amc}*\vt_b)\Big)\\
&&-\Big(\d (d\vp)-\vp_m\d\vt^m\Big)\wedge *d\vp+\d \vt^m \wedge (e_m\hook L).
\ern
Extract the total derivatives in the corresponding terms to obtain
 \br\label{4.15}
\d L &=&
\d\vt_{m}\wedge\Big( d(*\frac 12F^{mbc}\vt_{bc})+
\frac 12F^{abc}(C_{amb}*\vt_c-C_{amc}*\vt_b)+\vp_m \wedge *d\vp +e_m\hook L\Big)\nonumber\\ 
&&
+\d\vp \ d*d\vp 
+d\Big(\frac 12\d\vt_{a}\wedge*F^{abc}\vt_{bc}-\d\vp *d\vp\Big).
\er
The variation relation (\ref{4.15}) will play a basic role in the sequel. 
Let us rewrite it in a compact form by introducing the 
following abbreviated notations. \\
Define one-indexed 2-forms
 \begin{equation}\label{4.16}
\C^a:=\frac 12 C^{abc}\vt_{bc}=d\vt^a.
\end{equation}
and a conjugate strength 2-form 
 \begin{equation}\label{4.16a}
\F^a:=\frac 12 F^{abc}\vt_{bc}=(\rho_1+\rho_3)\C^a+
\rho_2e^a\hook(\vt^m\wedge\C_m)-\rho_3\vt^a\wedge(e_m\hook \C^m)
\end{equation}
The 2-form $\F^a$ can be also represented via the irreducible (under the 
Lorentz group) decomposition of the 2-form $\C^a$ (see \cite{Hehl98}). 
Write 
\begin{equation}\label{4.17a}
\C^a={}^{(1)}\C^a+{}^{(2)}\C^a+{}^{(3)}\C^a,
\end{equation}
where
\br\label{4.17b}
{}^{(1)}\C^a&=&\C^a-{}^{(2)}\C^a-{}^{(3)}\C^a,\nonumber\\
{}^{(2)}\C^a&=&\frac 13 \vt^a\wedge(e_m\hook \C^m),\nonumber\\
{}^{(3)}\C^a&=&\frac 13 e^a\hook(\vt_m\wedge \C^m).
\er
Substitute (\ref{4.17b}) into (\ref{4.16a}) to obtain 
\begin{equation}\label{4.17c}
\F^a=(\rho_1+\rho_3){}^{(1)}\C^a+(\rho_1-2\rho_3){}^{(2)}\C^a+
(\rho_1+3\rho_2+\rho_3){}^{(3)}\C^a.
\end{equation}
The coefficients in (\ref{4.17c}) coincide with those calculated in 
\cite{Hehl98}. \\ 
The 2-forms  $\C^a$ and $\F^a$ do not depend on a choice of 
a coordinate system. They  change as 
vectors by $SO(1,3)$ transformations of the coframe. \\
Using (\ref{4.16}) the coframe Lagrangian can be rewritten  as
 \begin{equation}\label{4.17}
{}^{(c)}L=\frac 14 C_{abc}F^{abc}*1=\frac 12 \C_a\wedge *\F^a
 \end{equation}
Let us turn now to the variation relation (\ref{4.15}). 
The terms of the form $F\cdot C$  can be rewritten as 
\brn
&&F^{abc}(C_{amb}*\vt_c-C_{amc}*\vt_b)=(F^{abc}-F^{acb})C_{amb}*\vt_c
\nonumber\\
&&\qquad =2 C_{amb}*(e^b\hook \F^a)=-2(e_m\hook \C_a)\wedge *\F^a.
\ern
Consequently (\ref{4.15}) takes the form 
 \br\label{4.18}
\d L&=& \d\vt_m\wedge \Big(d(*\F^m)-(e_m\hook \C_a)\wedge *\F^a+
(e_m\hook d\vp)\wedge *d\vp+e_m\hook L\Big)\nonumber \\
&& \qquad \qquad +\d\vp \ d*d \vp+d(\d\vt_m\wedge \F^m-\d\vp*d\vp)
\er
Collect now the quadratic terms in the brackets in two different 3-forms. 
The  scalar field 3-form is defined as 
 \begin{equation}\label{4.19}
{}^{(s)}\T_m=-(e_m\hook d\vp)\wedge *d\vp +\frac 12 e_m\hook (d\vp\wedge*\vp),
 \end{equation}
while the coframe field 3-form is defined as
 \begin{equation}\label{4.20}
{}^{(c)}\T_m=(e_m\hook \C_a)\wedge *\F^a -\frac 12 e_m\hook (\C_a\wedge *\F^a).
 \end{equation}
In (\ref{4.19}) and (\ref{4.20}) the explicit expressions for the  scalar and the coframe Lagrangians are 
inserted. We will use also the  notation
 \begin{equation}\label{4.21}
\T_m={}^{(s)}\T_m+{}^{(c)}\T_m
 \end{equation}
for the total 3-form of the system.\\
Using the definitions above  the variational relation (\ref{4.15}) 
results in the final form
  \begin{equation}\label{4.22}
\d L= \d\vt_m\wedge \Big(d*\F^m-\T^m\Big)+\d\vp \ d*d \vp+
d(\d\vt_m\wedge \F^m-\d\vp*d\vp).
\end{equation}
%%%%%%%%%%%%%%%%%%%%%%%%%%%%%%%%%%%%%%%%%%%%%%%%%%%%
 \subsection{The field equations}                 %%%%%%% 4
%%%%%%%%%%%%%%%%%%%%%%%%%%%%%%%%%%%%%%%%%%%%%%%%%%%%
We are ready now to write down the field equations.
For the scalar field it takes the familiar form
 \begin{equation}\label{4.23}
d*d\vp=0
\end{equation}
As for the coframe, the field equation is
 \begin{equation}\label{4.24} 
d*\F^m=\T^m.
\end{equation}
Note that this is the same equation as (\ref{3.6}). 
The equivalence of the forms can be established by straightforward but tedious 
algebraic manipulations. 
The form of the coframe field  equation (\ref{4.24}) is exactly the same 
as discussed above: 
exterior derivative of the dual strength in the left hand side and a 3-form 
in the right hand side. 
Thus the 3-forms $\T^m$ serves as a source for the strength $\F^m$. 
The field equation (\ref{4.24}) yields the conservation law
\begin{equation}\label{cons-I}
d\T_m=0.
\end{equation}
Thus we obtain a conserved total 3-form for the system which is 
constructed from the first order derivatives
of the field variables (coframe). 
It is local and covariant. 
Moreover, it is naturally reduced to the sum of a 
scalar field 3-form and a coframe field 3-form. 
The 3-form $\T_m$ is our  candidate for the coframe energy-momentum current. 
We obtained this expression
directly from the field equation and it is similar to the
electromagnetic current 3-form (\ref{2.7}).   
Note however an important distinction.
The source term for electromagnetic field  depends only on material field 
(scalar in our case). The electromagnetic field itself is uncharged. 
As for the the coframe field its source is the sum of a  material (scalar) 
field current and a coframe field  current ${}^{(c)}\T_m$. \\
In order to identify the conserved 3-form $\T_m$ with the 
energy-momentum current we have to answer the question: 
{\it What symmetry this conserved current can be associated with?}
%%%%%%%%%%%%%%%%%%%%%%%%%%%%%%%%%%%%%%%%%%%%%%%%%%%%
\subsection{Conserved currents}                       %%%%%%% 5
%%%%%%%%%%%%%%%%%%%%%%%%%%%%%%%%%%%%%%%%%%%%%%%%%%%%
Return to the variational relation (\ref{4.22}). 
On shell, for fields satisfying the field equations (\ref{4.23})
and (\ref{4.24}), it takes the form
 \begin{equation}\label{4.26}   
\d L=d \Theta(X),
\end{equation}
where
\begin{equation}\label{4.27}  %{var4-1}
\Theta=\d\vt^a\wedge *\F_a-\d\vp*d\vp. 
\end{equation}
Consider the variations of the fields produced by the Lie derivative taken 
relative to a smooth vector field $X$ on the manifold $M$.
The total Lagrangian (\ref{4.10}) is a diffeomorphic invariant, 
hence it's  variation is produced by the Lie  derivative taken relative 
to the same vector field $X$. 
The  Lie derivative of an arbitrary  4-form is a 
total derivative thus the the relation (\ref{4.26}) takes a 
form of a conservation law for a certain 3-form
\begin{equation}\label{4.28} 
dJ(X)=0, \qquad J(X)=\Theta(X)-X\hook L.
\end{equation}
The explicit form of this conserved current is 
 \begin{equation}\label{4.29} 
J(X)=\Big(d(X\hook\vt^a)+X\hook \C^a \Big)\wedge *\F_a-(X\hook d\vp)*d\vp
-X\hook L.
\end{equation}
As in the case of electromagnetic field,
this  quantity includes a term (the first one) which is non-linear 
(non-algebraic) relative to an arbitrary 
vector field $X$. 
Such a  form of the conserved current can not be 
used for definition of an energy-momentum tensor. 
Unlike the electromagnetic case this problem can be solved 
merely by using the canonical form of the current.\\ 
Let us  take $X=e_a$. The first term of  (\ref{4.29})  vanishes  
identically.  Thus  
\begin{equation}\label{4.30} 
J(e_m)=(e_m\hook \C^a )\wedge *\F_a-(e_m\hook d\vp)*d\vp-e_m\hook L.
\end{equation}
Observe that this expression coincides with the source of the field equation 
(\ref{4.24}): $J(e_m)=\T_m$. 
Thus the conserved current $\T_m$ is associated with the diffeomorphic 
invariance of the Lagrangian and consequently represents 
the energy-momentum current.\\
Another way to avoid the first term in (\ref{4.29}) is to extract the 
total derivative. 
\brn
J(X)&=&d\Big((X\hook\vt^a)\wedge *\F_a\Big)-(X\hook\vt^a)\wedge d*\F_a
+(X\hook \C^a) \wedge *\F_a\\
&&-(X\hook d\vp)*d\vp-X\hook L.\\
&=&d\Big((X\hook\vt^a)*\F_a\Big)-(X\hook\vt^a) (d*\F_a-\T_a) 
\ern
Thus, up to the field equation (\ref{4.24}), the current $J(X)$ represents  
a total derivative of a certain 2-form 
  \begin{equation}\label{4.31a} 
J(X)=d\Big((X\hook\vt^a)*\F_a\Big).
\end{equation}
This result is a special case of a general proposition due 
to Wald \cite{Wald}   for  a diffeomorphic Lagrangians. 
The  2-form  
\begin{equation}\label{4.31} 
{}^{(c)}Q(X)=(X\hook \vt^a) *\F_a.
\end{equation}
can be  referred to as the {\it Noether charge for the coframe field}. 
The canonical form of the Noether charge for the coframe field 
coincides with the dual of the conjugate strength $\F^a$. 
\begin{equation}\label{4.32} 
{}^{(c)}Q_a={}^{(c)}Q(e_a)=*\F_a.
\end{equation}
Note, that such  2-form plays an important role in Wald's treatment 
of the black hole entropy \cite{Wald}.
%%%%%%%%%%%%%%%%%%%%%%%%%%%%%%%%%%%%%%%%%%%%%%%%%%%%
\subsection{Energy-momentum tensor}                       %%%%%%% 5
%%%%%%%%%%%%%%%%%%%%%%%%%%%%%%%%%%%%%%%%%%%%%%%%%%%%
In this section we present the expressions for the energy-momentum tensor 
for the scalar and coframe fields.   
Apply the definition (\ref{2.28}) to the conserved current
(\ref{4.19}).
Thus, the energy-momentum tensor for the scalar field  is
\begin{equation}\label{4.38}
{}^{(s)}T_{mn}=*\Big((e_m\hook d\vp)\wedge *(e_n\hook d\vp)\Big)+
\frac 12 \eta_{mn}*(d\vp\wedge*d\vp).
\end{equation}
Observe that this expression is  symmetric 
${}^{(s)}T_{mn}={}^{(s)}T_{nm}$ and 
coincides with the familiar coordinate-wise expression. \\
As for the coframe field, its energy-momentum tensor is derived 
from the current (\ref{4.20})
\begin{equation}\label{4.39}
{}^{(c)}T_{mn}=e_n\hook*{}^({c})\T_m=e_n\hook*\Big((e_m\hook\C_a)\wedge*\F^a-
\frac 12 e_m\hook(\C_a\wedge*\F^a)\Big).
\end{equation}
Using (\ref{A.15}) we rewrite the first term in (\ref{4.39}) as 
\brn
&&e_n\hook*\Big((e_m\hook\C_a)\wedge*\F^a\Big)=
-*\Big(\vt_n\wedge*^2\Big[(e_m\hook\C_a)\wedge*\F^a\Big]\Big)\\
&& \ \ \ =*\Big((e_m\hook\C_a)\wedge*^2(\vt_n\wedge*\F^a)\Big)=
-*\Big((e_m\hook\C_a)\wedge*(e_n\hook\F^a)\Big).
\ern
As for the second term in (\ref{4.39}) it takes the form 
\brn
&&-\frac 12 e_n\hook*\Big(e_m\hook(\C_a\wedge*\F^a)\Big)=
\frac 12*\Big(\vt_n\wedge *^2(e_m\hook(\C_a\wedge*\F^a)\Big)\\
&&=-\frac 12*(\vt_n\hook*\vt_m)*(\C_a\wedge*\F^a)=
\frac 12\eta_{mn}*(\C_a\wedge*\F^a).
\ern
Consequently the energy-momentum tensor for the coframe field is  
\begin{equation}\label{4.40}
{}^{(c)}T_{mn}=-*\Big((e_m\hook\C_a)\wedge*(e_n\hook\F^a)\Big)+
\frac 12\eta_{mn}*(\C_a\wedge*\F^a).
\end{equation}
Observe that this expression is exactly of the same form as the familiar 
expression for the energy momentum tensor of the Maxwell 
electromagnetic field (\ref{2.34a}). 
It  also satisfies the following \\
{\bf Proposition}
{\it For all teleparallel models described by the Lagrangian (\ref{4.1}) 
the energy-momentum tensor defined by 
(\ref{4.40}) is traceless $ {}^{(c)}{T^m}_m=0.$}\\
{\bf Proof}
Compute the trace of (\ref{4.40}): 
\brn
&&{}^{(c)}{T^m}_m={}^{(c)}T_{mn}\eta^{mn}=-*\Big((e_m\hook\C_a)\wedge*(e^m\hook\F^a)\Big)+2*(\C_a\wedge*\F^a)\\
&&=*\Big((e_m\hook\C_a)\wedge*^2(\vt^m\wedge*\F^a)\Big)+2*(\C_a\wedge*\F^a)\\
&&=-*\Big(\vt^m\wedge(e_m\hook\C_a)\wedge*\F^a\Big)+2*(\C_a\wedge*\F^a)=0
\ern
In the latter equality the relation  (\ref{A.9}) was used. 
%%%%%%%%%%%%%%%%%%%%%%%%%%%%%%%%%%%%%%%%%%%%%%%%%%%%
 \subsection{The field equation for a general system}           %%%%%%% 5
%%%%%%%%%%%%%%%%%%%%%%%%%%%%%%%%%%%%%%%%%%%%%%%%%%%%
The coframe field equation have been derived for a coframe-scalar 
system. Consider a general minimally coupled system which includes 
a coframe field $\vt^a$ and a material field $\psi$. 
The material field can be  a differential form of an arbitrary degree 
 and can carry exterior and interior indices. 
Take the total Lagrangian of the system to be in the form 
\begin{equation}\label{4-41}
L={}^{(c)}L(\vt^a,d\vt^a)+{}^{(m)}L(\vt^a,\psi, d\psi),
\end{equation}
where the coframe Lagrangian is defined by (\ref{4.1}). 
Take the variation of (\ref{4-41}) relative to the coframe field $\vt^a$ 
\begin{equation}\label{4-42}
\d L=\d\vt_a\wedge\Big(d*\F^a-{}^{(c)}\T^a-{}^{(m)}\T^a\Big),
\end{equation}
where the material form current 3-form is defined via the variation 
derivative of the material Lagrangian relative to the coframe field $\vt^a$
\begin{equation}\label{4-42a}
{}^{(m)}\T^a:=-\frac {\d}{\d\vt_a}{}^{(m)}L.
\end{equation}
Consequently, the field equation for the general system (\ref{4-41}) takes the 
form 
\begin{equation}\label{4-43}
d*\F^a=\T^a,
\end{equation}
where the notion of the total current of the system 
$\T={}^{(c)}\T^a+{}^{(m)}\T^a$ 
is introduced. 
In the tensorial form the equation can be rewritten as 
\begin{equation}\label{4-44}
e_b\hook *d*\F^a={T^a}_b,
\end{equation}
or equivalently 
\begin{equation}\label{4-45}
\vt_b\wedge d*\F^a={T^a}_b*1.
\end{equation}
The conservation law for the total current $d\T^a=0$ is a consequence of the 
field equation (\ref{4-43}). \\
The field equation is similar to the Maxwell field equation for the 
electromagnetic field. 
Observe, however, an important difference. 
The source term in the right hand side of the electromagnetic field equation 
depends only on external fields. 
The electromagnetic field itself is not charged and does not produces 
additional fields.\\
On the other hand the tensorial form (\ref{4-44}) of the field equation 
is similar to the 
Einstein field equation for the metric tensor
\begin{equation}\label{4-46}
{R^a}_b-\frac 12 \d^a_b R={{}^{(m)}T^a}_b.
\end{equation}
Again, the source terms in the field equations (\ref{4-44}) and  (\ref{4-46}) 
are  different. 
The source of the Einstein gravity is the energy-momentum tensor  only of the  
materials fields. 
The conservation of this tensor is a consequence of the field equation. 
Thus even if  some meaningful conserved energy-momentum current for 
the metric field 
existed  it would have been conserved regardless of  the material field current. 
Consequently, any transmission of the energy-momentum current between the 
material 
and gravitational fields is  forbidden in the framework of the traditional 
Einstein gravity.  \\
As for the coframe field equation, the total energy-momentum current plays 
a role of the source of the field. 
Consequently the coframe field is completely ``self-interacted'' - the 
energy-momentum current of the coframe field produces an additional field. 
The conserved current of the coframe field equation is the total 
energy-momentum current, not only the material current. 
Thus the transmission of the current between  the material field and the 
coframe field is, in principle, possible. 
%%%%%%%%%%%%%%%%%%%%%%%%%%%%%%%%%%%%%%%%%%%%%%%%%%%%
\section{Non-gravity teleparallel models}          %%%%%%% 5
%%%%%%%%%%%%%%%%%%%%%%%%%%%%%%%%%%%%%%%%%%%%%%%%%%%%
Let us turn now to concrete teleparallel models, which can be constructed 
by choosing a fixed set of parameters $\rho_i$  in the Lagrangian (\ref{4.1}). 
We start with  two models, which have unique static, 
 spherical symmetric solutions. 
These solutions, however, have not  Newtonian limit at infinity. 
So these models are not viable models  of gravity. 
We continue with the so-called teleparallel equivalent of gravity. 
This model has a unique  static, 
 spherical symmetric solution, which leads to the Schwarzschild metric. 
The consideration is restricted to the vacuum case. 
%%%%%%%%%%%%%%%%%%%%%%%%%%%%%%%%%%%%%%%%%%%%%%%%%%%%
\subsection{Yang-Mills-type model}     %%%%%%% 5
%%%%%%%%%%%%%%%%%%%%%%%%%%%%%%%%%%%%%%%%%%%%%%%%%%%%
We start with a teleparallel model described by a Lagrangian 4-form  
of a  pure Yang-Mills type
\begin{equation}\label{5-1}
L=\frac 12 \rho d\vt^a\wedge *d\vt_a.
\end{equation}
Here the choice of the free parameters in the Lagrangian (\ref{4.1}) is 
\begin{equation}\label{5-2}
\rho_1=\rho, \qquad \rho_2=\rho_3=0.
\end{equation}
The conjugate strength (\ref{4.16a}) takes the form 
\begin{equation}\label{5-3}
\F^{a}=\rho \ \C^{a}.
\end{equation}
Consequently the field equation (\ref{4-43}) is 
\begin{equation}\label{5-4}
d(* \ \C^a)=\frac 1{\rho}\T^a,
\end{equation}
where the source term is the energy-momentum current of the coframe field 
\begin{equation}\label{5-5}
\T^a=\rho\Big((e^a\hook  \ \C_m)\wedge * \ \C^m-\frac 12 e^a\hook ( \ \C_m\wedge * \ \C^m)\Big).
\end{equation}
It was proved in \cite{it5} that 
the field equation (\ref{5-4}) has a unique
static solution of a "diagonal" form, which is spherical symmetric 
($r=\sqrt{x^2+y^2+z^2}$)
\begin{equation}\label{5-6}
\vt^0=\Big(\frac {r_0}r \Big)^{4/3}dt, 
\qquad \vt^\a=\Big(\frac {r_0}r \Big)^{2/3}dx^\a \qquad \a=1,2,3.
\end{equation}
The corresponding metric is 
\begin{equation}\label{5-7}
ds^2=\Big(\frac {r_0}r \Big)^{8/3}dt^2-
\Big(\frac {r_0}r \Big)^{4/3}(dx^2+dy^2+dz^2).
\end{equation}
Substituting the solution (\ref{5-6}) in (\ref{5-5}) we obtain 
\begin{equation}\label{5-8}
\T_0=-\frac 49\frac{\rho}{{r_0}^2}\Big(\frac {r_0}r\Big)^{2/3}*\vt_0,
\end{equation}
\begin{equation}\label{5-9}
\T_\alpha=\frac 49\frac{\rho}{{r_0}^2}\Big(\frac {r_0}r\Big)^{2/3}\Big(-2\d_{\a\b}+5\frac{x_\a x_\b}{r^2}\Big)*\vt^\b,
\end{equation}
where $\a,\b=1,2,3$ and $\d_{\a\b}=diag\{1,1,1\}$.\\
The energy-momentum tensor for the coframe field takes the form
\begin{equation}\label{5-10}
T_{mn}=-\frac 49\frac{\rho}{{r_0}^2}\Big(\frac {r_0}r\Big)^{2/3}
\left(\begin{array}{cccc}
1&0&0&0\\
0&-2+5\frac{x^2}{r^2}&5\frac{xy}{r^2}&5\frac{xz}{r^2}\\
0&5\frac{xy}{r^2}&-2+5\frac{y^2}{r^2}&5\frac{yz}{r^2}\\
0&5\frac{xz}{r^2}&5\frac{yz}{r^2}&-2+5\frac{y^2}{r^2}
\end{array}\right).
\end{equation}
Observe that this matrix  is traceless and symmetric. The leading 
coefficient in (\ref{5-6}) can be interpreted as follows. 
Compute the scalar curvature of the metric (\ref{5-7}). It  is negative 
at every point of $M$ and coincides with the leading coefficient of 
(\ref {5-10}) 
\begin{equation}\label{5-11}
R=-\frac 49 \frac 1{r_0^2}\Big(\frac {r_0}r \Big)^{2/3}.
\end{equation} 
Identify the energy of the field (\ref{5-6}) with the $\{00\}$ 
component of the energy-momentum tensor to obtain
\begin{equation}\label{5-12}
E=-\frac 49\frac{\rho}{{r_0}^2}\Big(\frac {r_0}r\Big)^{2/3}.
\end{equation}
Thus, in order to have positive energy of the field, we have to require 
that the parameter $\rho$ should be negative. 
The choice $\rho=-1$ is in accordance with \cite{Hehl98}. \\
Observe a remarkable  relation between the energy of the coframe field  
and the scalar curvature
\begin{equation}\label{5-13}
E=\rho R.
\end{equation}
It should be noted, however, that the right hand side of (\ref{5-13}) is 
invariant under local $SO(1,3)$ transformations of the coframe as well as 
under arbitrary diffeomorphisms of the manifold. The left hand side of 
(\ref{5-13}), however, depends on a specific  coframe and on a choice 
of the coordinates. \\  
It was proved in \cite{it5} that the solution of the type 
(\ref{5-9}) is a  generic solution in a wide class of 
teleparallel models. 
For almost arbitrary choice of the parameters $\rho_i$ the unique 
spherical symmetric static solution has a  form 
 \begin{equation}\label{5-14}
\vt^0=\Big(\frac {r_0}r \Big)^a dt, 
\qquad \vt^i=\Big(\frac {r_0}r \Big)^bdx^i,
\end{equation}
where the exponents $a$ and $b$ are functions of $\rho_i$ 's 
\begin{equation}\label{5-14a}
a=\frac{2\r_1}{2\r_3-3\r_1}, \qquad b=\frac{\r_1-2\r_3}{2\r_3-3\r_1}.
\end{equation}
The corresponding metrics is 
 \begin{equation}\label{5-21}
ds^2=\Big(\frac {r_0}r \Big)^{2a} dt^2-\Big(\frac {r_0}r \Big)^{2b}
(dx^2+dy^2+dz^2).
\end{equation}
The metric (\ref{5-21}) is 
singular at the origin of the coordinates as well as at infinity 
and regular for $0<r<\infty$. 
Thus it does not have the Minkowskian limit and can not be used 
to describe the gravitational field.
%%%%%%%%%%%%%%%%%%%%%%%Figure 1:  %%%%%%%%%%%%%%%%
%
% Change the graph 4/9  !!!!
%
\begin{figure}[t]
\label{fig:1}
%\begin{center}
%%\leavevmode
\epsfxsize=2.25in
%%\epsfbox{k=3.ps}
\includegraphics[angle=270,width=8cm]{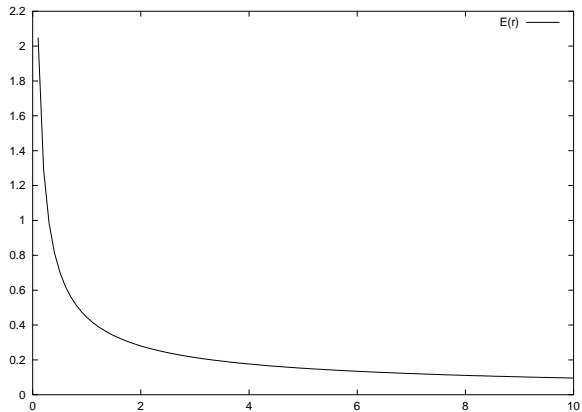}
%\end{center}
\caption{
The energy (\ref{5-12}) plotted as a function of a distance.
}
\end{figure}
%%%%%%%%%%%%%%%%%%%end Figure 1:  k=3  %%%%%%%%%%%%%%%%

%%%%%%%%%%%%%%%%%%%%%%%%%%%%%%%%%%%%%%%%%%%%%%%%%%%%
 \subsection{Anti Yang-Mills-type model.}          %%%%%%% 5
%%%%%%%%%%%%%%%%%%%%%%%%%%%%%%%%%%%%%%%%%%%%%%%%%%%%
Consider a teleparallel model with a Lagrangian of the form
\begin{equation}\label{5-22}
L=\rho d^\dagger\vt^a\wedge *d^\dagger\vt_a.
\end{equation}
This is  obtained from the Yang-Mills Lagrangian (\ref{5-1}) by replacing the 
familiar exterior derivative operator with the coderivative operator 
$d^\dagger=*d*$. 
The coderivative of the basis 1-forms can be expressed by 
its exterior derivative \cite{hehl95}
\begin{equation}\label{5-23}
d^\dagger\vt^a=*(\vt^{am}\wedge *d\vt_m)=e^a\hook(e_m \hook \d\vt^m).
\end{equation}
Consequently, by using the rule (\ref{A.9}), the Lagrangian can be rewritten as
\begin{equation}\label{5-24}
L=-\rho  \Big(e_a\hook(e_m \hook d\vt^m)\Big)\wedge  (\vt^{an}\wedge *d\vt_n)=
\rho (e_m \hook d\vt^m)\wedge \vt^{n}\wedge *d\vt_n. 
\end{equation}
In order to embed the Lagrangian (\ref{5-24}) into the general form of the 
teleparallel Lagrangian (\ref{4.1}) we use  again the rule   
(\ref{A.9}) to obtain
\brn
&&L=\rho (e_m \hook (d\vt^m\wedge \vt^{n})-d\vt^n)\wedge *d\vt_n\\
&&=\rho d\vt_n\wedge \vt_{m}\wedge*(d\vt^m\wedge \vt^{n})-
\rho d\vt^n\wedge *d\vt_n%=-\rho {}^{(1)}L+\rho {}^{(3)}L
\ern
Thus the Lagrangian (\ref{5-22}) is
\begin{equation}\label{5-22a}
L=-\rho \  {}^{(1)}L+\rho \  {}^{(3)}L,
\end{equation}
and the corresponding set of parameters is 
\begin{equation}\label{5-24a}
\rho_1=-\rho,\qquad \rho_2=0,\qquad \rho_3=\rho.
\end{equation}
The conjugate strength is
\begin{equation}\label{5-25}
\F^a:=\rho\vt^a\wedge(e_m\hook \C^m)
\end{equation}
The field equation (\ref{4.24}) takes the form 
\begin{equation}\label{5-26}
 d*\Big(\vt^a\wedge(e_m\hook \C^m)\Big)=\frac 1\rho \T^a,
\end{equation}
where the energy-momentum current $\T^a$ is obtained from (\ref{4.20}) by 
inserting the conjugate strength (\ref{5-25}).\\
A unique  spherical symmetric, static, ``diagonal'' 
solution for the equation (\ref{5-26}) is
\begin{equation}\label{5-24b}
\vt^0=dt,\qquad \vt^i=\Big(1+\frac {r_0}{r}\Big)dx^i.
\end{equation}
This coframe generates an asymptotically-flat metric
\begin{equation}\label{5-29}
ds^2=dt^2-\Big(1+\frac {r_0}{r}\Big)^2(dx^2+dy^2+dz^2).
\end{equation}
The metric (\ref{5-29}) represents a point-like solution if the 
corresponding ADM-mass is positive. 
Rewrite the metric (\ref{5-29})  in the spherical  
Schwarzschild-type coordinates.
In the spherical coordinates the metric is 
$$ds^2=dt^2-\Big(1+\frac {r_0}{r}\Big)^2(dr^2+r^2d\Omega^2).$$
Using now the translation
$$\tilde{r}= r+r_0$$ we obtain the asymptotic-flat 
metric in the Schwarzschild coordinates
\begin{equation}\label{6-18}
ds^2=dt^2-\frac {d\tilde{r}^2}{(1-\frac {r_0}{\tilde{r}})^2}-
\tilde{r}^2d\Omega^2.
\end{equation}
The ADM mass for the metric (\ref{6-18}) takes the form
\begin{equation}\label{6-19}
m:=\lim_{\tilde{r}\to \infty}\frac {\tilde r}2\Big(1-(1-
\frac {r_0}{\tilde{r}})^2\Big)=r_0.
\end{equation}
Thus by taking the parameter $r_0$ to be positive we obtain a 
particle-type solution with a finite positive ADM-mass. \\
The metric (\ref{5-29}) is singular at the origin 
$r=0$ and consequently the metric (\ref{6-18}) is singular at 
$\tilde{r}=r_0$. 
In order to clarify the nature of this singularity compute the 
scalar curvature of the metric (\ref{5-29}) via the formula (\ref{6-1}). 
The result is  
\begin{equation}\label{6-20}
R=\frac {r_0^2}{(r_0+r)^4}=\frac {r_0^2}{\tilde r^4}.
\end{equation}
This function is non-zero and regular for all  values of $r$ 
including the origin ($r=0$). \\
The proper distance for a radial null geodesic in the metric 
(\ref{6-18}) is equal to the proper time and attach the infinity
$$
l=t=\int^{\tilde r_1}_{r_0} \frac {d\tilde r}{1-\frac {r_0}{\tilde r}} 
\to \infty
$$
Thus the point $r=0$ ($\tilde r=r_0$) does not belong to any 
final part of the space-time. \\
Computing the conjugate strength with the coframe (\ref{5-24}) we obtain 
\begin{equation}\label{5-29a}
\F^a=\frac{2m\rho}{r(m+r)^2}\d_{\mu\nu}x^\mu\vt^{a\nu},
\end{equation}
where $a=0,1,2,3$, while $\mu,\nu=1,2,3$.\\
Calculating the energy-momentum current for the coframe field (\ref{5-24}) we 
obtain
\begin{equation}\label{5-30}
\T^0=\frac{2m^2\rho}{(m+r)^4}*\vt^{0},
\end{equation}
\begin{equation}\label{5-31}
\T^\a=-\frac{2m^2\rho}{(m+r)^4}\frac{x^\a x^\mu}{r^2}\d_{\mu\nu}*\vt^{\nu},
\end{equation}
Consequently the energy-momentum tensor obtains the form 
\begin{equation}\label{5-32a}
T_{mn}=\frac{2m^2\rho}{(m+r)^4}
\left(\begin{array}{cccc}
1&0&0&0\\
0& -\frac{x^2}{r^2}&-\frac{xy}{2r^2}&-\frac{xz}{2r^2}\\
0& -\frac{xy}{2r^2}&-\frac{y^2}{r^2}&-\frac{yz}{2r^2}\\
0& -\frac{xz}{2r^2}&-\frac{yz}{2r^2}&-\frac{z^2}{r^2}
\end{array}\right).
\end{equation}
Observe that this matrix  is traceless and symmetric. The leading 
coefficient in (\ref{5-6}) is proportional again to the scalar curvature. 
The free coefficient $\rho$ have to be chosen to be positive. Taking 
$\rho=\frac 12$ we obtain also in this model the relation $E=R$. 
%%%%%%%%%%%%%%%%%%%%%%%Figure 3 %%%%%%%%%%%%%%%%
%
% 
\begin{figure}[t]
\label{fig:2}
%\begin{center}
%%\leavevmode
\epsfxsize=2.25in
%%\epsfbox{k=3.ps}
\includegraphics[angle=270,width=8cm]{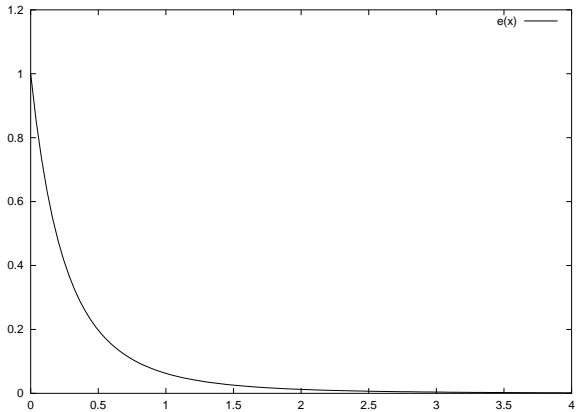}
%\end{center}
\caption{The energy (\ref{5-32a}) plotted as a function of a distance.
}
\end{figure}
%%%%%%%%%%%%%%%%%%%end Figure 1:  k=3  %%%%%%%%%%%%%%%%
%%%%%%%%%%%%%%%%%%%%%%%%%%%%%%%%%%%%%%%%%%%%%%%%%%%%
\section{Teleparallel models for gravity}          %%%%%%% 5
%%%%%%%%%%%%%%%%%%%%%%%%%%%%%%%%%%%%%%%%%%%%%%%%%%%%
Let us delete the first (Yang-Mills-type) term in the general 
Lagrangian (\ref{3.2}) by requiring 
$$ \rho_1=0.$$
Thus the Lagrangian under consideration is 
\begin{equation}\label{6-1}
L=\rho_2(d\vt_a \wedge \vt^a ) \wedge*(d\vt_b\wedge\vt^b)+
\rho_3 (d\vt_a \wedge\vt^b ) \wedge *(d\vt_b \wedge \vt^a ).
\end{equation}
The conjugate strength (\ref{4.16a}) takes the form
\begin{equation}\label{6-2}
\F^a=\rho_2e^a\hook(\vt^m\wedge\C_m)+
\rho_3 e^m\hook (\vt^a\wedge\C_m)
\end{equation}  
The field equation (\ref{4.24}) takes the form 
\begin{equation}\label{6-3}
 \rho_2 d\Big(*(\vt^m\wedge\C_m)\wedge\vt^a\Big)+
\rho_3 d\Big(* (\vt^a\wedge\C_m)\wedge\vt^m\Big)=\T^a,
\end{equation}
where the energy-momentum current $\T^a$ is obtained from (\ref{4.20}) by 
inserting the conjugate strength (\ref{5-25}).\\
A unique  spherical symmetric, static, ``diagonal'' 
solution for the equation (\ref{6-3}) is
\br\label{6-4}   
\vt^0&=&\frac{1-m/{2r}}{1+ m/{2r}}dt,\qquad \vt^i=
\Big(1+\frac m{2r}\Big)^2dx^i, \qquad i=1,2,3,
\er
This coframe corresponds to the Schwarzschild metric in 
 isotropic coordinates 
\begin{equation}\label{5-29b}
ds^2=\Big(\frac{1-m/{2r}}{1+ m/{2r}}\Big)^2dt^2-\Big(1+\frac m{2r}\Big)^4
(dx^2+dy^2+dz^2).
\end{equation}
Observe that the first term of the strength (\ref{6-2}) as well as the first 
term of the field equation (\ref{6-3}) vanish identically for an arbitrary 
choice of the ``diagonal''  form $\vt^a=Fdx^a$.\\
Computing the conjugate strength for the coframe (\ref{5-24}) we obtain 
\begin{equation}\label{5-29c}
\F^0=\frac{-2m\rho_3}{r^3(1+m/{2r})^3}\d_{\mu\nu}x^\mu\vt^{0\nu},
\end{equation}
\begin{equation}\label{5-29d}
\F^\a=\frac{-2m\rho_3(1-m/{4r})}{r^3(1-m/{2r})(1+m/{2r})^3}\d_{\mu\nu}x^\mu\vt^{\a\nu},
\end{equation}
where $a=0,1,2,3$, while $\a,\mu,\nu=1,2,3$.\\
Calculating the energy-momentum current for the coframe field (\ref{5-24}) we 
obtain
\begin{equation}\label{5-30a}
\T^0=\frac{-3\rho_3(m^2/r^4)(1-m/{6r})}{(1-m/{2r})(1+m/{2r})^6}*\vt^{0},
\end{equation}
\begin{equation}\label{5-31a}
\T^\a=\frac {\rho_3({m^2}/{r^4})}{(1-m/{2r})(1+m/{2r})^6}*\Big(\vt^\a-
\frac{m}{r}\frac{x^\a x^\mu}{r^2}\d_{\mu\nu}\vt^{\nu}\Big),
\end{equation}
Consequently the energy-momentum tensor is
\begin{equation}\label{5-32}
T_{mn}=\frac {\rho_3({m^2}/{r^4})}{(1-m/{2r})(1+m/{2r})^6}
\left(\begin{array}{cccc}
3-\frac m{2r}&0&0&0\\
0& 1-\frac m{2r}\frac {x^2}{r^2}&-\frac m{2r}\frac{xy}{r^2}&-\frac m{2r}\frac{xz}{r^2}\\
0& -\frac m{2r}\frac{xy}{r^2}&1-\frac m{2r}\frac{y^2}{r^2}&-\frac m{2r}\frac{yz}{r^2}\\
0& -\frac m{2r}\frac{xz}{r^2}&-\frac m{2r}\frac{yz}{r^2}&1-\frac m{2r}\frac{z^2}{r^2}
\end{array}\right).
\end{equation}
Observe that this matrix  is traceless and symmetric. The leading 
coefficient in (\ref{5-6}) is proportional again to the scalar curvature. 
The free coefficient $\rho$ has to be positive.
Taking 
$\rho=\frac 12$ we obtain also in this model the relation $E=R$. 
%%%%%%%%%%%%%%%%%%%%%%%Figure 1:  k=3 %%%%%%%%%%%%%%%%
%
% 
\begin{figure}[t]
\label{fig:3}
%\begin{center}
%%\leavevmode
\epsfxsize=2.25in
%%\epsfbox{k=3.ps}
\includegraphics[angle=270,width=8cm]{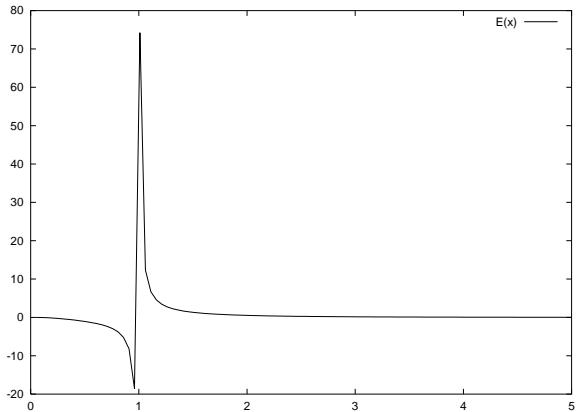}
%\end{center}
\caption{The energy (\ref{5-32}) plotted as a function of a distance.
}
\end{figure}
\section*{Acknowledgments}
I am deeply grateful to F.W. Hehl and to S. Kaniel for constant support, interesting discussions and valuable comments.

\appendix

%%%%%%%%%%%%%%%%%%%%%%%%%%%%%%%%%%%%%%%
\section{Basic notations and definitions}
%%%%%%%%%%%%%%%%%%%%%%%%%%%%%%%%%%%%%%%%
Let us list our basic conventions. 
We consider an $n$-dimensional differential manifold $M$
of signature
\begin{equation}\label{A.1}
\eta_{ab}=diag(-1,+1,\cdots,+1).
\end{equation}
Let the manifold $M$ will be endowed with a smooth coframe field
(1-forms) 
\begin{equation}\label{A.2}
\{\vt^a(x), \ a=0,\cdots,n-1\}.
\end{equation}
Note that a smooth non-degenerate frame (coframe) field can be
defined on a manifold of a zero second Stiefel-Whitney class.
However this topological restriction is not exactly relevant
in physics because the solutions of physical field equations
can degenerate at a point or on a curve.
Moreover, these solutions produce the most important
physical models (particles, strings, etc.).\\
The coframe $\vt^a(x)$ represents, at a given point $x\in M$, a basis
of the linear space of 1-forms $\Omega^1$.
The set of all non-zero exterior products 
of basis 1-forms
\begin{equation}\label{A.3}
\vt^{a_1,\cdots,a_p}:=\vt^{a_1}\wedge\cdots\wedge \vt^{a_p}
\end{equation}
represents a basis of the linear space of $p$-forms $\Omega^p$.
Note the (anti)commutative rule for arbitrary forms $\a\in \Omega^p$
and $\b\in \Omega^q$
\begin{equation}\label{A.4}
\a\wedge\b=(-1)^{pq}\b\wedge\a.
\end{equation}
The dual set of vector fields
\begin{equation}\label{A.5}
\{e_a(x), \ a=0,\cdots,n-1\}
\end{equation}
forms a basis of the linear space of vector fields at a given point.\\
The duality of vectors and 1-forms can be expressed by {\it inter
product} operation for which we use the symbol $\hook$. Namely,
\begin{equation}\label{A.6}
e_a\hook\vt^b=\d^b_a.
\end{equation}
The action $X\hook w$ of  a vector $X$  on a form $w$ 
of arbitrary degree $p$ is defined by requiring: 
(i) linearity in $X$ and in $w$,  
(ii) modified Leibniz rule for the wedge product of
$\a\in \Omega^p$  and $\b \in \Omega^q$
\begin{equation}\label{A.7}
X\hook (\a\wedge\b)=(X\hook\a)\wedge\b+(-1)^p\a\wedge(X\hook\b).
\end{equation}
These properties together with (\ref{A.6}) guarantee the uniqueness of the
map $\hook:\Omega^p \to \Omega^{p-1}$.\\
The following relations involving the inner product operation ($p=deg(w)$)
are useful for actual calculations.
\br\label{A.8}
X\hook(Y\hook w)&=&-Y\hook(X\hook w),\\
\label{A.9}
\vt^a\wedge (e_a\hook w)&=&pw,\\
\label{A.10}
e_a\hook(\vt^a\wedge w)&=&(n-p)w.
\er
We use also the forms $\vt_a:=\eta_{ab}\vt^b$ with subscript and the
corresponding vector fields $e^a:=\eta^{ab}e_b$ with superscript. Thus
\begin{equation}\label{A.11}
e_a\hook\vt_b=\eta_{ab}.
\end{equation}
The linear spaces $\Omega^p$ and $\Omega^{n-p}$ have the same dimensions
${n\choose k}={n\choose{n-p}}$. Thus they are isomorphic.
This isomorphism {\it Hodge dual map} is linear.
Thus it is enough
to define its action on basis forms:
\begin{equation}\label{A.12}
*(\vt^{a_1\cdots a_p})=\frac 1{(n-p)!}
\epsilon^{a_1\cdots a_pa_{p+1}\cdots a_n}\vt_{a_{p+1}\cdots a_n}.
\end{equation}
We use here the complete antisymmetric pseudo-tensor
$\epsilon^{a_1\cdots a_{n-1}}$ normalized as $\epsilon^{01\cdots (n-1)}=1$.
The set of indices $\{a_1,\cdots,a_n\}$ is an even permutation of the
standard set $\{0,1,\cdots,(n-1)\}$.\\
Thus $*\vt^{0\cdots (n-1)}=1$ and $*1=-\vt^{0\cdots (n-1)}$.\\
The consequence of the definition (\ref{A.12}) is ($deg(\a)=deg(\b)$)
\begin{equation}\label{A.13}
\a\wedge *\b=\b\wedge\a.
\end{equation}
For the choice of the signature (\ref{A.1}) we obtain
\begin{equation}\label{A.14}
*^2w=(-1)^{p(n-p)+1}w.
\end{equation}
In the case $n=4$ the operator $*^2$ preserves the forms of  odd degree
and changes the sign  of the forms of  even degree.\\
The following equation is useful for actual calculations
\begin{equation}\label{A.15}
e_a\hook w=-*(\vt_a\wedge *w).
\end{equation}
To prove this linear relation it is enough to check it for
the basis forms.\\
The pseudo-orthonormality for the basis forms
$\vt^a$ yields the 
{\it metric tensor} $g$ on the manifold $M$ 
\begin{equation}\label{A.16}
g=\eta_{ab}\vt^a\otimes\vt^b.
\end{equation}   
The formulas (\ref{A.11}) and (\ref{A.15}) can be applied to derive
a useful form of a scalar product of two vectors $X$ and $Y$.
We write these vectors in the basis $e_a$ as $X=X^me_m$ and $Y=Y^me_m$.
Thus the scalar  product is
$$<X,Y>=X^mY^n<e_m,e_n>=X^mY^n\eta_{mn}.$$
Using (\ref{A.11}) we obtain
$$<X,Y>=X^mY^n(e_m\hook\vt_n)$$
Thus
\begin{equation}\label{A.17}  %%%% possible \flat
<X,Y>=X\hook {}^\sharp Y=Y\hook {}^\sharp X,
\end{equation}
where ${}^\sharp X$ is the 1-form dual to the vector $X$ which obtained by 
a canonical map  from vectors to 1-forms
$$\sharp:X^me_m\to X^m\vt_m.$$

\end{document}